\documentclass[aps,twocolumn,showpacs]{revtex4}
\usepackage{epsfig}
\usepackage{amsbsy}
\usepackage{amsmath}
\usepackage{amsfonts}
\usepackage{graphicx}
\usepackage{latexsym}

\def\ket#1{|#1\rangle}

\newcommand{\be}{\begin{equation}}
\newcommand{\ee}{\end{equation}}
\newcommand{\rf}[1] {(\ref{#1})}
\newcommand{\nn}{\nonumber}

\begin{document}

\title{Entangling power and operator entanglement in qudit systems}
\author{Xiaoguang Wang, Barry C Sanders, and Dominic W Berry}
\affiliation{Department of Physics and Centre for Quantum Computer Technology, \\
Macquarie University, Sydney, New South Wales 2109, Australia.}
\date{\today}

\begin{abstract}
We establish the entangling power of a unitary operator on a general
finite-dimensional bipartite quantum system with and without ancillas, and give relations between the entangling power based on 
the von Neumann entropy and the entangling power based on the linear entropy. 
Significantly, we demonstrate that the entangling power of a general
controlled unitary operator acting on two equal-dimensional qudits is
proportional to the corresponding operator entanglement if linear
entropy is adopted as the quantity representing the degree of entanglement. 
We discuss the entangling power and operator entanglement of
three representative quantum gates on qudits: the SUM, double SUM,
and SWAP gates.
\end{abstract}
\maketitle

\section{Introduction}
Entanglement has been established as a crucial resource for quantum
information tasks such as quantum communication and quantum computation
\cite{Nie00}. Consequently, generating or enhancing entanglement between
separate physical systems is of paramount importance in quantum information
theory, and two cases are typically studied: (i) ancilla-assisted
entanglement generation, and (ii) entanglement generation without assistance
from ancillas. Significant effort is currently directed to quantifying
entanglement of states; similarly it is important to quantify entanglement
capabilities \cite{Zan00,Dur,Mak00,Lei02,Kra01,Ben02,Chi02,Ber02,Wol02}
of unitary operations, or more generally, the ``strength'' \cite{Nie02}
of the operator.

Entangling power based on the linear entropy~\cite{Zan00} is a valuable, and relatively easy to calculate,
measure of the entanglement capability of an operator. We extend this
definition to the ancilla-assisted case, and establish an equivalence between
entangling power and an alternative quantity, ``operator entanglement''
\cite{Zan01,Wan02}, for arbitrary controlled unitary operations acting on two
equal-dimensional qudits.

Any gate that creates entanglement between qudits without ancillas acts as
a universal gate for quantum computation when assisted by arbitrary one-qudit
gates \cite{Bry02,Bre02}. Therefore, the SUM gate \cite{Got98,Alb00,Bar02,San02}
[a generalization of the controlled-NOT (CNOT) gate for qubits] can be chosen
as the basic, or primitive, two-qudit gate for qudit-based quantum computation.
We study entangling power of the SUM gate and other two-qudit gates, namely the double-SUM (DSUM) and SWAP gate to illustrate our
results on more general gates as well as the general applicability of our
approach.

This paper is organized as follows. In Sec.\ \ref{Sec:measures}, we introduce
the von Neumann entropy and linear entropy as entanglement measures. 
In Sec.\ \ref{Sec:powers}, we review the entangling power based on the linear entropy without ancillas, and
extend to entangling power assisted by ancillas. We also give relations between the entangling power based on the von Neumann entropy and the entangling power based on the linear entropy. In Sec.\ \ref{Sec:control}, we
study the entangling power of a general controlled unitary operaror $C_U$, and
build an equivalence relation between entangling power and operator
entanglement. We also provide an example of $C_U$ resulting from higher-order
spin-spin interactions. In Sec.\ \ref{Sec:Ent}, we discuss entangling
capabilities of representative two-qudit gates, including SUM
\cite{Got98,Alb00,Bar02,San02}, DSUM and SWAP gates, and summarize our results
in Sec.\ \ref{Sec:conc}.

\section{Entanglement measures}
\label{Sec:measures}
Various measures of entanglement exist, each with its own advantages and
disadvantages \cite{Kra01}. Two commonly used entanglement measures for
pure states are the von Neumann entropy $\tilde{E}$ and the linear entropy $E$.
For a two-qudit pure state $|\Psi\rangle\in{\cal H}_{d}\otimes{\cal H}_{d}$ they
are defined as
\begin{align}
\tilde{E}(\ket{\Psi})&:=-\text{Tr}_1[\rho_1 \ln \rho_1]\label{e1},\\
E(|\Psi \rangle )&:=\text{Tr}_1[\rho_1(1-\rho_1)]=1-\text{Tr}_1\rho_1 ^2,
\label{e2}
\end{align}
where $\rho_1 =\text{Tr}_2(|\Psi \rangle\langle \Psi |)$ is the reduced
density matrix. For convenience, we use natural logarithms throughout this paper.
The von Neumann entropy that we define therefore differs from the usual von
Neumann entropy by a factor of $\ln 2$.
The von Neumann entropy and the linear entropy satisfy the inequalities
\begin{equation}
0\leq\tilde{E}(|\Psi \rangle )\leq \ln d,\quad 
0\leq E(|\Psi \rangle )\leq 1-1/d,
\end{equation}
where the lower (upper) bound is reached if and only if $|\Psi\rangle$
is a product state (maximally entangled state). 

The entanglement measures discussed above can also be applied to the study of
entanglement of operators \cite{Zan01}.
An operator can increase entanglement of a state, but an operator can also be
considered to be entangled because operators themselves inhabit a Hilbert space.
The entanglement of quantum operators is introduced \cite{Zan01} by noting that
the linear operators over ${\cal H}_d$ span a $d^2$-dimensional Hilbert space
with the scalar product between two operators $X$ and $Y$ given by the
Hilbert-Schmidt product $\langle X,Y\rangle:=\text{Tr}(X^{\dagger }Y)$, and
$||X||_{\rm HS}:=\sqrt{\text{Tr}(X^{\dagger }X)}$. We denote this
$d^2$-dimensional Hilbert space as ${\cal H}_{d^2}^{\text{HS}}$. Thus, the
operator acting on ${\cal H}_{d_1}\otimes {\cal H}_{d_2}$ is a state in the
composite Hilbert space ${\cal H}_{d_1^2}^{\text{HS}}\otimes
{\cal H}_{d_2^2}^{\text{HS}}$, and the entanglement of an operator
$X$ is well-defined \cite{Zan01}.

Any operator $O$ (not necessarily unitary) acting on ${\cal H}_{d_1}\otimes
{\cal H}_{d_2}$ may be Schmidt-decomposed as \cite{Nie02}
$
O=\sum_ns_nA_n\otimes B_n,
$
where $s_n\geq 0$ and $\{A_n\}$ and $\{B_n\}$ are {orthonormal} operator
bases for systems $1$ and $2$. From the Schmidt form, entanglement measures
for a unitary operator $U$ can be determined to be
\begin{align}
\tilde{E}(U)&=-\sum_n\frac{s_n^2}{d_1d_2}\ln\left(\frac{s_n^2}{d_1d_2}\right),
\\ E(U)&=1-\frac 1{d_1^2d_2^2}\sum_ns_n^4,
\end{align}
where the factor $1/(d_1d_2)$ arises from normalization of the unitary operator.

\section{Assisted and unassisted entangling powers}
\label{Sec:powers}
The entangling power of a unitary operator $U$ is defined over
${\cal H}_{d_1}\otimes {\cal H}_{d_2}$ as the average entanglement of the
state $U|\psi_1\rangle\otimes|\psi_2\rangle$ for product states
$|\psi_1\rangle\otimes|\psi_2\rangle\in {\cal H}_{d_1}\otimes {\cal H}_{d_2}$.
The entangling power $\tilde{e}_{\text{p}}(U)$ based on the von Neumann
entropy and $e_{\text{p}}(U)$ based on the linear entropy are given by
\cite{Zan00,Symbol}
\begin{align}
\tilde{e}_{\text{p}}(U)&=\int d\mu(\psi_1,\psi_2)\tilde{E}(U|\psi_1\rangle
\otimes |\psi_2\rangle ), \label{epuepu}\\
e_{\text{p}}(U)&=\int d\mu(\psi_1,\psi_2)E(U|\psi_1\rangle \otimes |\psi_2
\rangle ), \label{epu}
\end{align}
where $d\mu(\psi_1,\psi_2)$ denotes an integral measure over product states.

These two entangling powers are related, and relations between 
linear entropy and von Neumann entropy have been investigated~\cite{Wei1,Berry1}.
Let us first
rewrite Eqs.\ (\ref{epuepu}) and (\ref{epu}) in the form
\begin{align}
\tilde{e}_{\text{p}}(U)&=\int d\mu(\psi_1,\psi_2) \sum_i \big[
-\lambda_i(\psi_1,\psi_2) \ln \lambda_i(\psi_1,\psi_2) \big], \\
e_{\text{p}}(U)&=1-\int d\mu(\psi_1,\psi_2)\sum_i\lambda_i(\psi_1,\psi_2)^2,
\end{align}
where $\lambda_i(\psi_1,\psi_2)$ are the squares of the coefficients in the
Schmidt decomposition of $U|\psi_1\rangle \otimes |\psi_2\rangle$. In addition,
let us define the entangling power
\begin{equation}
\bar e_{\text{p}}(U)=-\ln (1-e_{\text{p}}(U)), \label{powerp}
\end{equation}
which is a monotonic function of the entangling power $e_{\text{p}}(U)$, and satisfies $\bar e_{\text p}(U)\ge e_{\text p}(U)$.

Evaluating $\tilde e_{\text{p}}(U)-\bar e_{\text{p}}(U)$ gives
\begin{align}
&\tilde e_{\text{p}}(U)-\bar e_{\text{p}}(U) \nn \\
&=\int d\mu(\psi_1,\psi_2) \sum_i
\left[-\lambda_i(\psi_1,\psi_2) \ln \frac{\lambda_i(\psi_1,\psi_2)}
{1-e_{\text{p}}(U)} \right] \nn \\ &\ge \int d\mu(\psi_1,\psi_2) \sum_i
\lambda_i(\psi_1,\psi_2) \left[1-\frac{\lambda_i(\psi_1,\psi_2)}
{1-e_{\text{p}}(U)} \right] \nn \\ &= 1-\frac 1{1-e_{\text{p}}(U)}
\int d\mu(\psi_1,\psi_2) \sum_i\lambda_i(\psi_1,\psi_2)^2 \nn \\ &= 0.
\end{align}
This result implies that 
$
\tilde{e}_{\text{p}}(U)\ge \bar{e}_{\text{p}}(U). 
$
Another useful bound on $\tilde{e}_{\text{p}}(U)$ can be obtained by noting that
the average entanglement generation cannot be larger than the maximum
entanglement generation:
$
\tilde e_{\rm p}(U) \le \tilde E_{\rm max}(U),
$
where
\begin{equation}
\tilde E_{\rm max}(U) = \max_{|\psi_1\rangle,|\psi_2\rangle}
\tilde{E}(U|\psi_1\rangle\otimes |\psi_2\rangle ).
\end{equation}
These relations are useful because $\bar e_{\text{p}}(U)$ and $\tilde E_ {\rm max}(U)$ may be determined
analytically, and used to draw conclusions about $\tilde e_{\text{p}}(U)$.

Now we investigate the entangling power based on the linear entropy.
The calculation of linear entropy $E$ can be simplified by doubling the Hilbert
space from ${\cal H}_{d_1}\otimes {\cal H}_{d_2}$ to ${\cal H}
_{d_1}\otimes {\cal H}_{d_2}\otimes{\cal H}_{d_1}\otimes {\cal H}_{d_2}$ and
using the identity $\text{Tr}_{12}[(\hat{A}\otimes \hat{B})S_{12}]=\text{Tr}_1
(\hat{A}\hat{B})$ \cite{Zan00}.
Here $S_{ij}$ denotes the swap  operation between equal-dimensional systems $i$
and $j$.
It is clear from Eq.\ (\ref{epu}) that different integral measures give
different entangling powers. For the Haar measure, group theory techniques yield
\cite{Zan00}
\begin{align}
e_{\text{p}}(U)&=1-\frac 1{d_1(d_1+1)d_2(d_2+1)}\big[ d_1d_2^2+d_2d_1^2
\nn \\
&~~~+\text{Tr}_{1234}(U^{\otimes 2}S_{13}U^{\dagger \otimes 2}S_{13})
\nn \\
&~~~+\text{Tr}_{1234}(U^{\otimes 2}S_{24}U^{\dagger \otimes
2}S_{13}) \big].  \label{epu1}
\end{align}

This definition of entangling power presents an anomaly that the entangling
power of a SWAP gate over a $d\times d$ space is zero \cite{Zan00}. The
entangling power defined by Eq.\ (\ref{epu}) does not include the advantage of
incorporating ancilla assistance. With assistance from ancillas the SWAP gate
can generate entanglement. The dimension of each ancilla can be chosen as the
dimension of the original system because the Schmidt number of a state in the
composite system of the original system plus ancilla is at most the dimension of
the original system \cite{Nie02}.

Let the SWAP gate for systems $A$ and $B$ act on the state
$|\Psi \rangle_{A^{\prime}A}\otimes|\Phi\rangle_{BB^{\prime}}$, where 
\begin{align}
|\Psi \rangle _{A^{\prime }A}	&=\sum_{n=0}^{d-1}|n\rangle _{A^{\prime
}}\otimes |n\rangle_A \, \in {\cal H}_{d}\otimes {\cal H}_{d},\\
|\Phi \rangle _{BB^{\prime }} &=\sum_{n=0}^{d-1}|n\rangle _B\otimes
|n\rangle _{B^{\prime }}\, \in {\cal H}_{d}\otimes {\cal H}_{d},
\end{align}
and $A^{\prime }$ and $B^{\prime }$ denote ancillas for $A$ and $B$,
respectively. The final state after applying $U$ will have entanglement
increased by $E=1-1/d^2$. The entanglement increase draws on the ancillary
resources. Without these ancillas, the SWAP gate cannot increase entanglement,
which is the case considered in Ref.\ \cite{Zan00}.

We consider a $d_1\times d_2$ system, and introduce two ancillas $A'$ and $B'$
with dimension $d_1$ and $d_2$, respectively. Then, the whole state space
expands to ${\cal H}_{d_1}^{\otimes 2}\otimes{\cal H}_{d_2}^{\otimes 2}$ in
which the first and fourth systems are ancillas. Let the unitary operator $U$
act on the whole state space ${\cal H}_{d_1}^{\otimes 2}\otimes
{\cal H}_{d_2}^{\otimes 2}$. Analogous to Eq.\ (\ref{epu}), we define the
ancilla-assisted entangling power as
\begin{equation}
e_{\text{p}}^{\text{anc}}(U)=\int d\mu(\alpha,\beta) E(U|\alpha\rangle_{12}
\otimes |\beta \rangle_{34}), \label{epuuu}
\end{equation}
where $|\alpha\rangle_{12}\in {\cal H}_{d_1}^{\otimes 2}$ and 
$|\beta\rangle_{34}\in {\cal H}_{d_2}^{\otimes 2}$. 
By splitting the whole system as subsystems 12 and 34, extending
Eq.\ (\ref{epu1}), and using the Haar measure, we obtain the assisted entangling
power as
\begin{align}
e_{\text{p}}^{\text{anc}}(U) &= 1-\frac 1{d_1^2(d_1^2+1)d_2^2(d_2^2+1)}
\{d_1^2d_2^4+d_2^2d_1^4 \nn \\
&~~~+\text{Tr}_{12\ldots 8}[U^{\otimes 2}(S_{15}S_{26})U^{\dagger \otimes
2}(S_{15}S_{26})]  \nn \\
&~~~+\text{Tr}_{12\ldots 8}[U^{\otimes 2}(S_{37}S_{48})U^{\dagger \otimes
2}(S_{37}S_{48})]\},  \label{eqau1}
\end{align}
where the state space now involved has doubled to 
${\cal H}_{d_1}^{\otimes 2}\otimes {\cal H}_{d_2}^{\otimes 2}
\otimes {\cal H}_{d_1}^{\otimes 2}\otimes {\cal H}_{d_2}^{\otimes 2}$.
Operator $S_{ij}S_{kl}$ is the swap between systems $i$ and $k$ and systems $j$
and $l$. We are interested only in the case that the unitary operator acts on
the system and not the ancillas, i.e., 
$
U\equiv I_{14}\otimes U_{23}. 
$
Equation (\ref{eqau1}) enables the calculation of the assisted entangling power
of $U$.

Qudit quantum computation is normally considered for many qudits with equal
dimension \cite{Bry02}. We will mainly examine the entangling powers of
two-qudit quantum gates as building blocks of the quantum computer, and
therefore we restrict to the case of equal dimension \cite{Lei02} ($d_1=d_2$).
In this case it is found that the entangling power of a unitary operator $U$ is
related to the entanglement of quantum unitary operators \cite{Zan01,Wan02}. 
The operator entanglement of unitary operator $U$ is given by \cite{Zan01}
\begin{equation}
E(U)=1-\frac 1{d^4}\text{Tr(}U^{\otimes 2}S_{13}U^{\dagger \otimes
2}S_{13}),  \label{eq:e}
\end{equation}
where $1/d^4$ is just the normalization factor for $U^{\otimes 2}$. From
Eqs.\ (\ref{epu1}) and (\ref{eq:e}), it is straightforward to verify
\cite{Zan00}
\begin{equation}
e_{\text{p}}(U)=\left(\frac{d}{d+1}\right)^2\left[ E(U)
+E(US_{12})-E(S_{12})\right].
\label{ep}
\end{equation}
Thus, the unassisted entangling power defined on $d\times d$ systems can be
expressed in terms of the entanglement of three operators, $U$, $U S_{12}$, and
$S_{12}$. Therefore, by studying the entanglement of these three operators we
can determine the entangling power of $U$.

From Eqs.\ (\ref{eqau1}) and (\ref{eq:e}), a similar result can be
obtained for the assisted entangling power as follows
\begin{align}
e_{\text{p}}^{\text{anc}}(U)&=\left(\frac{d^2}{d^2+1}\right)^2\nn\\
&~~~\times[E(U)+E(US_{13}S_{24})-E(S_{13}S_{24})].  \label{epa}
\end{align}
Note that relations (\ref{ep}) and (\ref{epa}) hold only when we quantify the
entanglement by the linear entropy, and from these relations we know that
unassisted and assisted entangling powers are completely determined by the
operator entanglement of $U$, $US_{12}$, and $US_{13}S_{24}$ (the entangling
powers of $S_{12}$ and $S_{13}S_{24}$ are given below). Based on these results
for the entangling powers, we next investigate a general controlled-$U$ quantum
operation on qudits.

\section{A general two-qudit controlled-$U$ gate}
\label{Sec:control}
A general controlled-$U$ quantum operation on two qudits is given by
\begin{equation}
C_U:=\sum_{n=0}^{d-1}P_{n,n}\otimes U_n,  \label{cu}
\end{equation}
with $P_{n,n}:=|n\rangle \langle n|$. We also define $P_{n,m}:=|n\rangle
\langle m|$, which satisfies $P_{n,m}P_{k,l}=\delta_{mk}P_{n,l}$. The
controlled-$U$ gate implements the unitary operator $U_n$ on the second system
if and only if the first system is in the state $|n\rangle$. The unassisted
entangling power and operator entanglement have been computed for the $C_U$ with
$d$ orthogonal $U_n$ \cite{Zan00,Zan01}. Here, $U_n$ can be arbitrary unitary
operators.
For the controlled-$U$ operation we have the following proposition.

{\bf Proposition 1}:{\it \ For the general controlled-}$U${\it \ gate acting
on }${\cal H}_d\otimes {\cal H}_d$,  
\begin{align}
e_{\text{p}}(C_U) &=\left(\frac{d}{d+1}\right)^2E(C_U),  \label{epcu} \\
e_{\text{p}}^{\text{anc}}(C_U) &=\left(\frac{d^2}{d^2+1}\right)^2E(C_U).
\label{epacu}
\end{align}
Proof: From Eqs.\ (\ref{ep}) and (\ref{epa}), we only need to prove that 
\begin{equation}
E(C_US_{12})=E(S_{12}), \;
E(C_US_{13}S_{24})=E(S_{13}S_{24}).
\end{equation}

Let us first prove $E(C_US_{12})=E(S_{12})$. The swap operator $S_{12}$ can be
written as
\begin{equation}
S_{12}=\sum_{i=j=0}^{d-1}P_{i,j}\otimes P_{j,i}.  \label{s12}
\end{equation}
It is easy to check that $\langle P_{i,j},P_{k,l}\rangle =$Tr$%
(P_{j,i}P_{k,l})=\delta _{ik}\delta _{jl}$. Therefore, $S_{12}$ is in the
Schmidt form with Schmidt number $d^2$, and the operator entanglement is given by 
\begin{equation}
E(S_{12})=1-1/d^2.
\end{equation}

From Eqs.\ (\ref{cu}) and (\ref{s12}), we write the product of the
operators $C_U$ and $S_{12}$ as
\begin{equation}
C_US_{12} =\sum_{i,j}P_{i,j}\otimes U_iP_{j,i}.  \label{cus12}
\end{equation}

The operator product $C_US_{12}$ is also in the Schmidt form with Schmidt
number $d^2$ since
\begin{equation}
\langle U_iP_{j,i},U_kP_{l,k}\rangle =\text{Tr}(P_{i,j}U_i^{\dagger
}U_kP_{l,k}) 
=\delta _{ik}\delta _{jl}.
\end{equation}
Thus, the operator entanglement is
\begin{equation}
E(C_US_{12})=E(S_{12})=1-1/d^2, \label{es1}
\end{equation}
which complete the proof of Eq.\ (\ref{epcu}).

To prove Eq.\ (\ref{epacu}) we write
\begin{equation}
S_{13}S_{24}=\sum_{ijkl}(P_{i,j}\otimes P_{k,l})\otimes (P_{j,i}\otimes
P_{l,k}). 
\end{equation}
Then the operator product $C_US_{13}S_{24}$ is given by
\begin{equation}
C_US_{13}S_{24} 
=\sum_{ijkl}(P_{i,j}\otimes P_{k,l})\otimes (U_kP_{j,i}\otimes P_{l,k}).
\end{equation}
It is straightforward to check that 
operators $S_{13}S_{24}$ and $C_US_{13}S_{24}$ are in 
equivalent Schmidt forms; hence their entanglements are equal, i.e., 
\begin{equation}
E(C_US_{13}S_{24})=E(S_{13}S_{24})=1-1/d^4.  \label{e1324}
\end{equation}
This completes the proof of Eq.\ (\ref{epacu}). $\Box$

Proposition 1 builds an equivalence relation between entangling powers and
operator entanglement. The higher the operator entanglement, the higher the
entangling
powers are for the general $C_{U}$ gate. From Eqs.\ (\ref{epcu}) 
and (\ref{epacu}), we immediately find
\begin{equation}
\frac{e_{\text{p}}^{\text{anc}}(C_U)}{e_{\text{p}}(C_U)}=\left(\frac{d^2+d}
{d^2+1}\right)^2>1, 
\end{equation}
which means that the entangling power of $C_U$ is enhanced by introducing
ancillas. 

\begin{figure}
\includegraphics[width=0.40\textwidth]{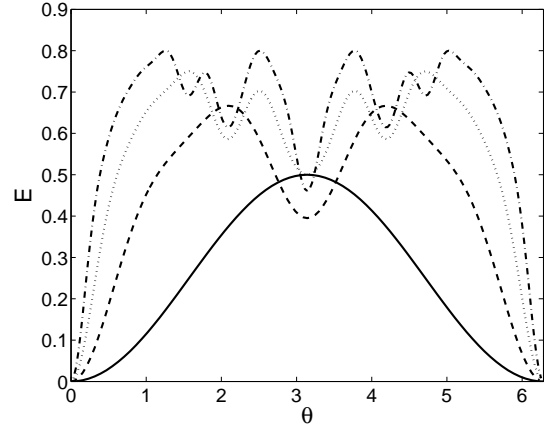}
\caption{\label{entfig}
The entanglement of the operator $U(\theta)$ vs $\theta$ for four different
spins: spin-1/2 (solid line), spin-1 (dashed line), spin-3/2 (dotted line), and spin-2 (dash-dot line).}
\end{figure}

We now apply Proposition 1 to study an example of a $C_{U}$ gate, and we will
see that the controlled-PHASE (CPHASE) gate \cite{Got98} is a special case of
this $C_U$ gate. We consider the interaction between two spin-$j$ systems via
the Hamiltonian \cite{WanSan00}
\begin{equation}
H=gJ_{1z}\otimes J_{2z}, 
\end{equation}
with $g$ the coupling strength and $J_{iz}$ the $z$-component of the
angular momentum operator $\vec{J}_i$. Up to local unitary operations, the
evolution operator $\exp (-igtJ_{1z}\otimes J_{2z})$ is equivalent to
$
U(\theta )=e^{i\theta N_1\otimes N_2}, 
$
where $N_i=J_{iz}+j$ and $\theta =-gt$. Note that the above unitary operator can
be written as
$
U(\theta )=\sum_{n=0}^{d-1}P_{n,n}\otimes e^{in\theta N_2}, 
$
where $d=2j+1$. Then the unitary operator $U(\theta )$ is a special case of the
controlled-$U$ gate.
The application of Proposition 1 to $U(\theta )$ tells us that the
entangling power $e_{\text{p}}$ and $e_{\text{p}}^{\text{anc}}$ are proportional
to the operator
entanglement of $U(\theta )$. Thus, we only need to calculate the operator
entanglement.

The unitary operator $U(\theta )$ can be rewritten as
\begin{equation}
U(\theta)=\sum_{m,n=0}^{d-1}\frac 1de^{i\theta mn}P_{n,n}\otimes
P_{m,m}, 
\end{equation}
where $1/d$ is just the normalization factor. We consider the
operators $U$ and $P_{n,n}$ as states $|U\rangle$ and $|P_{n,n}\rangle$,
where the bra-ket formalism is used. After tracing out the second system we
obtain the ``mixed operator'' for the first system,
\begin{equation}
\text{Tr}_2(|U\rangle \langle U|) =\sum_{mn}A_{mn}(\theta )|P_{m,m}\rangle
\langle P_{n,n}|,
\end{equation}
with
\begin{align}
A_{mn}(\theta ) &= \frac{1}{d^2} \sum_{k=0}^{d-1}e^{i\theta k(m-n)}
\nn \\
&=\frac{1}{d^2}\frac{\sin [d\theta (m-n)/2]}{\sin [\theta (m-n)/2]}e^{i(d-1)
\theta(m-n)/2}.  \label{trace}
\end{align}
For the case of two spin-1/2 systems it is straightforward to check that 
$E(U)=1/2\,\sin ^2(\theta /2)$ \cite{Zan01}. For higher spins we need to
find the eigenvalues of the $d\times d$ matrix $A$, from which the linear
entropy can be obtained. We numerically diagonalize the matrix, and the results
for the linear entropy are shown in Fig.\ \ref{entfig}.

From Fig.\ \ref{entfig}, we see that the entanglement is a periodic function of
$\theta $ with period $2\pi$, which can also be seen from Eq.\ (\ref{trace}).
The entanglement attains its maximum value of $1/2$ at $\theta=\pi$ for
spin-$1/2$, but does not reach its maximum value $1-1/d$ at $\theta=\pi$ for
spins greater than $1/2$. The first maximum value occurs at $\theta=2\pi/d$.
We also observe that there are two maximum values in one period for spin 1 and
spin 3/2 and four for spin 2. When $\theta =2\pi/d$ the unitary operator
$U(2\pi/d)$ becomes the CPHASE gate on qudits \cite{Got98}. Detailed analysis
of the operator entanglement for the CPHASE gate and other representative
quantum gates is provided in Sec.\ \ref{Sec:Ent}.

\section{Entanglement capability of qudit gates}
\label{Sec:Ent}
A qudit quantum computer is comprised of a network of one-qudit, two-qudit, and
multi-qudit gates. Two-qudit or multi-qudit gates usually have entanglement
capability. In this section, we calculate and compare entanglement capabilities
of different two-qudit gates. Before going to entangling gates let us first
review several useful one-qudit gates.

\subsection{One-Qudit gates}
Two essential one-qudit gates, denoted by $X$ and $Z$, are defined by their
action on the computational basis $|n\rangle\,(n=0,\ldots,d-1)$  
\begin{equation}
X|n\rangle =|n+1(\mathop{\rm mod}d)\rangle , \quad Z|n\rangle =\exp (i2\pi n/d)|n\rangle.
\end{equation}
Another useful quantum operation on qudits is the Fourier transformation $F$,
which is defined as
\begin{equation}
F|n\rangle =\sum_{k=0}^{d-1}\exp(i2nk\pi /d)|k\rangle .
\end{equation}
The Fourier transformation reduces to the Hadamard gate for the case of $d=2$.

\subsection{The CPHASE and SUM gate}
Henceforth we use $U_{\text{GATE}}$ to denote a two-qudit gate which includes
the CPHASE, SUM, DSUM, and SWAP gates. Now we examine the unitary operator
$U(\theta )=e^{i\theta {\cal N}_1\otimes {\cal N}_2}$ again. Note
that the number operator ${\cal N}$ is now simply defined as ${\cal N}|n\rangle
=n|n\rangle$. For $\theta=2\pi/d$, $U(\theta)$ can be written as
\begin{equation}
U_{\text{CPHASE}}=U(2\pi /d)=\sum_{n=0}^{d-1}P_{n,n}\otimes Z^n,  \label{phase}
\end{equation}
which is exactly the CPHASE gate \cite{Got98}. We will see that the CPHASE gate
differs from the SUM gate (defined below) only by local operations.

Let us consider one representative two-qudit gate, namely the SUM gate, which is
defined as \cite{Got98,Alb00,Bar02,San02}
\begin{equation}
U_{\text{SUM}}=U_{\text{SUM}}(1\rightarrow 2):=\sum_{n=0}^{d-1}P_{n,n}\otimes
X^n. \end{equation}
The notation $(1\rightarrow 2)$ indicates that the first qudit is the control
and the second qudit is the target. By using the Fourier transform we have
$
F^{-1}ZF=X.
$
Then, acting on the CPHASE gate $U_{\text{CPHASE}}$ by $I\otimes F^{-1}$
from the left and $I\otimes F$ from the right leads to the relation
between the CPHASE gate and the SUM gate,
\begin{equation}
U_{\text{SUM}}=(I\otimes F^{-1})U_{\text{CPHASE}}(I\otimes F).\label{relation}
\end{equation}

Relation \rf{relation} shows that the SUM and CPHASE gates differ only by local unitary operations. Therefore, they have same operator entanglement and entangling
powers. The SUM gate is an example of a general controlled-$U$ gate, and it has
Schmidt form
\begin{equation}
U_{\text{SUM}}=\sum_{n=0}^{d-1}\sqrt{d}\, P_{n,n}\otimes \big(X^n/\sqrt{d}\big).
\end{equation}
Thus, the entanglement of the SUM\ gate is given by 
\begin{equation}
E(U_{\text{SUM}})=1-1/d.
\end{equation}
According to Proposition 1 the unassisted and assisted entangling powers are
immediately evident. 

\subsection{The SWAP gate}
Another representative quantum gate is the SWAP gate $U_{\text{SWAP}}$,
which we have denoted by SWAP and $S_{ij}$ in the preceding sections.
We know that $E(U_{\text{SWAP}})=1-1/d^2$; now, from Eq.\ (\ref{ep}),
it is easy to ascertain that $e_{\text{p}}(U_{\text{SWAP}})=0$. 

Now we calculate the assisted entangling power of the SWAP gate. As we already
know the entanglement of operators $S_{23}$ and $S_{13}S_{24}$, only the
entanglement of the operator $S_{23}S_{13}S_{24}$ needs to be calculated.
The operator $S_{23}S_{13}S_{24}$ can be expressed as
\begin{align}
S_{23}S_{13}S_{24} &=\sum_{mnijkl}(P_{i,j}\otimes P_{m,n}P_{k,l})\otimes
(P_{n,m}P_{j,i}\otimes P_{l,k}) \nn\\
&=\sum_{il}{\cal P}_{i,l}\otimes {\cal P}_{i,l}^{\dagger },
\end{align}
with ${\cal P}_{i,l}=\sum_jP_{i,j}\otimes P_{j,l}$ 
satisfying the relations
\begin{equation}
\langle {\cal P}_{i,l},{\cal P}_{i^{^{\prime }},l^{^{\prime }}}\rangle
=d\delta _{ii^{\prime }}\delta _{ll^{\prime }}.
\end{equation}
Therefore, the Schmidt form of $S_{23}S_{13}S_{24}$ is given by
\begin{equation}
S_{23}S_{13}S_{24}=\sum_{il}d\,\big({\cal {P}}_{i,l}/\sqrt{d}\big)\otimes
\big({\cal {P}}_{i,l}^{\dagger}/\sqrt{d}\big), 
\end{equation}
from which the operator entanglement of $S_{23}S_{13}S_{24}$ is obtained as
\begin{equation}
E(S_{23}S_{13}S_{24})=E(S_{23})=1-1/d^2.
\end{equation} 
Then, substituting the above equation and Eq.\ (\ref{e1324})
into Eq.\ (\ref{epa}), we obtain
\begin{equation}
e_{\text{p}}^{\text{anc}}(S_{23})=\left( \frac{d^2-1}{d^2+1}\right) ^2. 
\end{equation}
After introducing ancillas the entangling power of the SWAP gate is no
longer zero. 

\subsection{Double SUM gate}
In this subsection we introduce and define a double SUM gate as
\begin{equation}
U_{{\text{DSUM}}}=U_{\text{SUM}}^{-1}(2\rightarrow 1)U_{\text{SUM}}
(1 \rightarrow 2)
\end{equation}
which can be considered as a generalization of the double CNOT gate
for qubits in the sense that the DSUM gate reduces to the double
controlled-NOT gate \cite{Zho00,Col01} for the case of dimension $d=2$.

Using the relation between SWAP and SUM gates given by \cite{Dab02,Swap1,Swap2}
\begin{align}
S_{12}&=(F^2\otimes I)U_{\text{SUM}}(1 \rightarrow 2)U_{\text{SUM}}^{-1}(2
\rightarrow 1)U_{\text{SUM}}(1 \rightarrow 2)  \nn \\
&=(F^2\otimes I)U_{\text{SUM}}U_{{\text{DSUM}}},
\label{ssum}
\end{align}
we observe that the SWAP gate can be constructed from three SUM gates and the
square of the Fourier transformation. This relation is useful for the following
analysis.

By using Eq.\ (\ref{ssum}) we can express DSUM as
\begin{equation}
U_{{\text{DSUM}}}
=U_{\text{SUM}}^{-1}(1 \rightarrow 2)S_{12}(I\otimes F^2),
\end{equation}
where the identity 
$
(\hat{A}\otimes \hat{B})S_{12}=S_{12}(\hat{B}\otimes \hat{A})
$
is used.
From Proposition 1, we know that for any controlled-$U$ we have
$E(C_US_{12})=E(S_{12})$. As $U_{\text{SUM}}^{-1}(1 \rightarrow 2)$ is a special
$C_U$ with $U=X^{-1}$, we have
\begin{align}
E(U_{{\text{DSUM}}}) &=E[U_{\text{SUM}}^{-1}(1 \rightarrow 2)S_{12}(I\otimes
F^2)] \nn\\ &=E[U_{\text{SUM}}^{-1}(1 \rightarrow 2)S_{12}]=E(S_{12}),
\end{align}
where the second equality is obtained by noticing that the local unitary
operators do not modify operator entanglement. Thus, we find that the
entanglement of the DSUM gate is equal to that of the SWAP gate. Using this fact
the unassisted entangling power is simplified to 
\begin{align}
e_{\text{p}}(U_{{\text{DSUM}}}) 
&=\frac{d^2}{(d+1)^2}E[U_{\text{SUM}}^{-1}(1 \rightarrow 2)S_{12}(I\otimes F^2)
S_{12}]  \nn \\
&=\frac{d^2}{(d+1)^2}E[U_{\text{SUM}}^{-1}(1 \rightarrow 2)]  \nn \\
&=e_{\text{p}}[U_{\text{SUM}}(1 \rightarrow 2)].
\end{align}
The last equality in the above equation results from the fact
$E(U)=E(U^\dagger)$ \cite{Zan01}. Therefore, the unassisted entangling power of
the DSUM gate is equal to that of the SUM gate.

To obtain the assisted entangling power of the DSUM gate we need to calculate
the entanglement of the operator $U_{{\text{DSUM}}}S_{13}S_{24}$. Up to local
unitary operations the operator is equivalent to $U_{\text{SUM}}^{-1}(1
\rightarrow 2)S_{23}S_{13}S_{24}$, which can be expressed as
\begin{align}
&U_{\text{SUM}}^{-1}(1 \rightarrow 2)S_{23}S_{13}S_{24}\nn\\
&=\sum_{iml}(P_{m,i}\otimes P_{i,l})\otimes
\Big(\sum_jP_{j-i,m}\otimes P_{l,j}\Big). 
\end{align}
It is straightforward to show the relations
\begin{align}
\langle P_{m,i}\otimes P_{i,l},P_{m^{\prime },i^{\prime }}\otimes P_{i^{\prime
},l^{\prime }}\rangle &=\delta _{mm^{\prime }}\delta _{ll^{\prime }}\delta
_{ii^{\prime }}, \nn \\
\left\langle \! \sum_jP_{j-i,m}\otimes P_{l,j},\sum_{j^{\prime }}P_{j^{\prime
}-i^{\prime },m^{\prime }}\otimes P_{l^{\prime },j^{\prime }}\!\!\right\rangle
&=d \delta _{mm^{\prime }}\delta _{ll^{\prime }}\delta _{ii^{\prime }}.
\end{align}
Thus, $U_{\text{SUM}}^{-1}(1 \rightarrow 2)S_{23}S_{13}S_{24}$ can be written in
the Schmidt form and
the entanglement 
\begin{equation}
E(U_{{\text{DSUM}}}S_{13}S_{24})=1-1/d^3
\end{equation}
follows.
Using the above equation, $E(U_{{\text{DSUM}}})=1-1/d^2$, and
$E(S_{13}S_{24})=1-1/d^4$, we obtain the assisted entangling power of the double
SUM gate as
\begin{equation}
e_{\text{p}}^{\text{anc}}(U_{{\text{DSUM}}})=\frac{d^4-d^2-d+1}{(d^2+1)^2}. 
\end{equation}

\begin{table}
\caption{Entangling powers ${e}_{\text{p}}$, ${e}_{\text{p}}^{\text{anc}}$, and
operator entanglement ${E}$ for the three representative two-qudit
gates.\label{linent}}
\begin{tabular}{cccccc}
Gates & $e_{\text{p}}$ & $e_{\text{p}}^{\text{anc}}$ & $E$ &  &  \\ \hline
$U_{\text{SUM}}$ & $\frac{d(d-1)}{(d+1)^2}$ & $\frac{d^3(d-1)}{(d^2+1)^2}$ &
$1-1/d$ &  &  \\ \hline
$U_{{\text{DSUM}}}$ & $\frac{d(d-1)}{(d+1)^2}$ & $\frac{d^4-d^2-d+1}{(d^2+1)^2}$
& $1-1/d^2$ &  &  \\ \hline
$U_{\text{SWAP}}$ & $0$ & $\frac{(d^2-1)^2}{(d^2+1)^2}$ & $1-1/d^2$ &  & \\
\hline
\end{tabular}
\end{table}

We summarize the results of the three representative quantum gates by Table
\ref{linent}. In particular, the two equalities
\begin{equation}
E(U_{\text{SWAP}})=E({U_{{\text{DSUM}}}}), \; e_{\text{p}}(U_{\text{SUM}})=
e_{\text{p}}({U_{{\text{DSUM}}}}),
\end{equation} 
hold.
As the entangling power $e_{\text{p}}$ of the DSUM gate is not zero, we can use
it as a universal gate in a qudit quantum computer.
Although the operator entanglement of the SWAP gate is equal to that of the DSUM
gate, we cannot use the SWAP gate as a universal gate since the corresponding
entangling power $e_{\text{p}}$ is zero. 

\subsection{Large dimension limit}
Now we consider the large $d$ limit. In this limit, we find that in every case
(except for the SWAP without ancillas) the values of $e_{\rm p}$ and
$e_{\text{p}}^{\text{anc}}$ approach 1. It is therefore better to consider the
measure $\bar e_{\rm p}$ \rf{powerp}, for the case without ancillas, and
$\bar{e}_{\text{p}}^{\text{anc}}=-\ln(1-e_{\text{p}}^{\text{anc}})$, for the
case with ancillas. The asymptotic expressions for these quantities for the
three different gates are given in Table \ref{asymp}.

\begin{table}
\caption{Asymptotic expressions for the entangling powers $\bar{e}_{\text{p}}$
and $\bar{e}_{\text{p}}^{\text{anc}}$, for the three representative two-qudit
gates. \label{asymp}}
\begin{tabular}{cccccc}
Gates & $\bar{e}_{\text{p}}$ & & $\bar{e}_{\text{p}}^{\text{anc}}$ &
&  \\ \hline
$U_{\text{SUM}}$ & $\ln d-\ln 3+O(d^{-1})$ & & $\ln d+O(d^{-1})$ & & \\ \hline
$U_{{\text{DSUM}}}$ & $\ln d-\ln 3+O(d^{-1})$ & & $2\ln d-\ln 3+O(d^{-1})$
& &\\ \hline
$U_{\text{SWAP}}$ & $0$ & & $2\ln d-\ln 4+O(d^{-2})$  & & \\ \hline
\end{tabular}
\end{table}

These asymptotic results may be used to gain information about the entangling
powers based on the von Neumann entropy, $\tilde e_{\rm p}$ and
$\tilde{e}_{\text{p}}^{\text{anc}}$. As was shown above,
$\tilde e_{\rm p}\ge\bar e_{\rm p}$, and it is also easily seen that
$\tilde e_{\rm p}^{\text{anc}}\ge\bar e_{\rm p}^{\text{anc}}$. In addition, the
maximum von Neumann entropy generation for each of the operations \cite{foot} is equal to the leading terms in the asymptotic expressions in Table \ref{asymp} .

These results allow us to accurately estimate the asymptotic values of
$\tilde e_{\rm p}$ and $\tilde{e}_{\text{p}}^{\text{anc}}$. For example, for the
case of the SUM gate without ancillas,
\begin{equation}
\ln d-\ln 3+O(d^{-1}) \le \tilde e_{\rm p}(U_{\text{SUM}}) \le \ln d.
\end{equation}
This result means that $\tilde e_{\rm p}(U_{\text{SUM}})=\ln d+O(1)$. The
corresponding results for the other cases are given in Table \ref{von}. In every
case, to leading order $\tilde{e}_{\text{p}}^{\text{anc}}$,
$\bar{e}_{\text{p}}^{\text{anc}}$, and the maximal entanglement are the same.

\begin{table}
\caption{Asymptotic expressions for the entangling powers based on the von
Neumann entropy, $\tilde{e}_{\text{p}}$ and $\tilde{e}_{\text{p}}^{\text{anc}}$,
for the three representative two-qudit gates. \label{von}}
\begin{tabular}{cccccc}
Gates & $\tilde{e}_{\text{p}}$ & & $\tilde{e}_{\text{p}}^{\text{anc}}$ &
&  \\ \hline
$U_{\text{SUM}}$ & $\ln d+O(1)$ & & $\ln d+O(d^{-1})$ & & \\ \hline
$U_{{\text{DSUM}}}$ & $\ln d+O(1)$ & & $2\ln d+O(1)$  & & \\ \hline
$U_{\text{SWAP}}$ & $0$ & & $2\ln d+O(1)$  & & \\ \hline
\end{tabular}
\end{table}

We therefore find that, in each of these cases (except the case of the
SWAP without ancillas)
\begin{equation}
\lim_{d\to\infty} \frac{\tilde{e}_{\text{p}}}
{\bar{e}_{\text{p}}} = \lim_{d\to\infty} \frac{\tilde{e}_{\text{p}}}
{\tilde{E}_{\text{max}}} = 1.
\end{equation}
In the case of the SUM gate with ancillas the agreement is particularly close.
Because the second term is of order $d^{-1}$, rather than order 1,
\begin{equation}
\lim_{d\to\infty} \tilde{e}_{\text{p}}^{\rm anc}(U_{\rm SUM})=\lim_{d\to\infty}
\bar{e}_{\text{p}}^{\rm anc} (U_{\rm SUM})=
\tilde{E}_{\text{max}}
(U_{\rm SUM}).
\end{equation}
That is, the average entanglement created approaches the maximum possible,
rather than just the ratio approaching 1. In addition, the results obtained for the SUM gate is applicable to any controlled-$U$ gate $C_U$ \rf{cu} with $d$ orthogonal $U_n$. 
\section{Conclusions}
\label{Sec:conc}
In conclusion, we have extended the entangling power based on the linear entropy
from the ancilla-unassisted case to the ancilla-assisted case. The assisted and
unassisted entangling powers, quantifying the average amount of entanglement
created by a unitary operator, turn out to be easy-to-use entanglement
capability measures which are complementary to the entanglement capability
measures based on the maximal entanglement \cite{Kra01,Ben02,Nie02} that an
operator can generate. 

We have studied the general controlled-$U$ operator and found that both the 
unassisted and assisted entangling powers are proportional to its operator
entanglement, which builds equivalence relations between the entangling power
and operator entanglement. This is important because the set of controlled-$U$
gates contains some very useful quantum gates such as the CPHASE and SUM gates,
and our result shows that it is sufficient to study the entanglement capability
by examining the operator entanglement. From the SUM gate, we have derived a new
quantum gate, the DSUM gate, which for qubits reduces to the double CNOT gate.
The entangling powers and operator entanglement of the SUM, DSUM, and SWAP gates
were examined in detail.

We have mainly considered the entangling power based upon the linear entropy. However, one is more interested in the entangling power based on the von Neumann entropy. 
Fortunately, the former provides a lower bound to the latter. In each of the cases we consider, our results show that for large dimension, to leading order the average entanglement created is equal to the maximum entanglement. Investigations of the entangling powers and operator entanglement will be
helpful in understanding the entangling capabilities of quantum operations as
physical resources, and will play an important role in quantum information
theory.

\acknowledgments
We acknowledge valuable discussions with Paolo Zanardi, Jamil Daboul, and Stephen D Bartlett.
This project has been supported by an Australian Research Council Large Grant and Macquarie University Research Fellowship.


\begin{thebibliography}{}
\bibitem{Symbol} In Ref.\ \cite{Zan00}, the entangling power was defined using a
different notation
$e_{\text{p}}(U):=\overline{E(U|\psi_1\rangle\otimes|\psi_2\rangle)}$.
\bibitem{foot}
The SUM gate may create entanglement of $\ln d$ from the state
$\sum_n |n\rangle|0\rangle$. This entanglement is the maximum because the SUM
gate may be implemented nonlocally using this much entanglement via a
generalization of the Gottesmann-Chuang \cite{gotchu} gate. Both the DSUM and
SWAP gates create entanglement of $2\ln d$ from initial states where the qudits
acted upon are maximally entangled with ancillas; this entanglement is the
maximum possible for a system of this dimension. Similarly, without ancillas
the maximum possible entanglement is $\ln d$, and this may be created via the
DSUM gate from the initial state $|0\rangle\sum_n|n\rangle$.
\end{thebibliography}
\end{document}